\documentstyle[12pt, epsfig]{article}
\begin{document}

\begin{flushright}
PM/99--49 \\
hep-ph/9912271
\end{flushright}
\begin{center}
{\Large \bf Analytical Study of Non-Universality of the Soft Terms
in the MSSM}
 \vspace{1cm}

{\sc D. Kazakov$^a$ and G. Moultaka$^b$}

 \vspace{0.5cm} {\it $^a$ Bogoliubov Laboratory of Theoretical
Physics, JINR, \\ 141 980 Dubna, Moscow Region,  Russia}

{\it$^b$ Physique Math\'ematique et Th\'eorique, UMR No 5825--CNRS, \\
Universit\'e Montpellier II, F--34095 Montpellier Cedex 5, France.
}
\end{center}
\vspace{1cm}

\begin{abstract}

We obtain general analytical forms for the solutions of the one-loop 
renormalization group equations in the top/bottom/$\tau$ sector of the MSSM. 
These solutions are valid for any value of $\tan \beta$ as well as any 
non-universal initial conditions for the soft SUSY breaking parameters 
and non-unification of the Yukawa couplings. We establish analytically a 
generic screening effect of non-universality, in the vicinity of the 
infrared quasi fixed point, which allows to determine sector-wise a hierarchy 
of sensitivity to initial conditions. We give also various numerical 
illustrations of this effect away from the quasi fixed point and assess 
the sensitivity of the Higgs and sfermion spectra to the non-universality 
of the various soft breaking sectors. As a by-product, a typical 
anomaly-mediated non-universality of the gaugino sector would have marginal 
influence on the scalar spectrum.

\end{abstract}
\newpage

\section{Introduction}

The Minimal Supersymmetric Standard Model (MSSM)  \cite{MSSM} has
been intensively studied recently \cite{many} with the emphasis on
prediction of  particle spectrum. It crucially depends on the
mechanism of SUSY breaking and the one which is commonly accepted
introduces the so-called  soft terms at a high energy scale
\cite{sugra, gaugemed}. These soft terms are running then down to
low energy according to the well known RG equations starting from
some initial values. In the minimal version the soft terms obey
the universality hypothesis which leaves one with a set of 5
independent parameters (before radiative electroweak symmetry breaking)
: $m_0, m_{1/2},A_0,\mu_0$ and $\tan\beta$.
However, recently there appeared some interest in new contributions
to SUSY breaking patterns which result in non-universal boundary
conditions for the soft terms \cite{Non}. To investigate the
influence of non-universality on the particle spectrum it is very
useful to have analytical solutions to the RG equations for the
evolution of the soft terms. Numerical analysis though
straightforward is rather complicated due to a large number of
free parameters while an analytical solution allows one to see
which of these parameters is important and which is not.

While with a single Yukawa coupling the analytical solution to the
one-loop RG equations has been known for long, for increasing
number of Yukawa couplings it has been obtained quite recently
\cite{AM} in the form that allows iterative presentation. On the
other hand, it has been shown \cite{K} that when one knows the
solution to RG equations for the couplings of a rigid theory, one
can obtain those for the soft terms by  a usual Taylor expansion
over the Grassmannian parameters. Thus one can apply the
Grassmannian expansion to the general solutions of \cite{AM} to
get the analytical solution for the soft terms with arbitrary
initial conditions in an iterative form and explore their
dependence on the latter.

In the present paper we perform this analysis in the MSSM with
three Yukawa couplings ($Y_t,Y_b$ and $Y_\tau$) and arbitrary
initial conditions for the soft terms in the one-loop
approximation. We examine the dependence of the solutions on the
initial conditions and show that if the Yukawa couplings are large
enough (typically bigger  than $\alpha_s$ at the GUT scale) the
solutions exhibit a quasi-fixed point behaviour. This means that in such
a regime some of the initial conditions are completely washed out, and
become actually inessential at low energy.
The main role is played by the gaugino masses
out of which the gluino mass is by far dominating.

\section{Analytic Solution to the RG Equations for Yukawa Couplings}

Though the RG equations for the Yukawa couplings do not have
explicit analytic solution, they can be solved iteratively as it
has been demonstrated in Ref.\cite{AM}. Using the notation
$$\alpha_i = \frac{g_i^2}{16\pi^2}, \ i=1,2,3; \ \  \ Y_k =
\frac{y_k^2}{16\pi^2}, \ k=t,b,\tau  $$
 one can write down the one-loop RG equations as
\begin{eqnarray}
\dot{\alpha}_i &=& -b_i\alpha_i^2, \label{a}\\ \dot{Y}_k &=&
Y_k(\sum_{i}c_{ki}\alpha_i - \sum_{l}a_{kl}Y_l), \label{y}
\end{eqnarray}
where  $\cdot \equiv d/dt, \ t= \log M_{GUT}^2/Q^2$ and
\begin{eqnarray*}
b_i&=&\{33/5,1,-3 \}, \\ c_{ti}&=& \{13/15,3,16/3 \}, \ \
c_{bi}=\{7/15,3,1 6/3 \},\ \  c_{\tau i}=\{9/5,3,0 \}, \\
a_{tl}&=&\{6,1,0 \},\ \ a_{bl}=\{1,6,1 \},\ \ a_{\tau l}=\{0,3,4
\}.
\end{eqnarray*}

The general solution to eqs.(\ref{a},\ref{y}) can be written as
\begin{eqnarray}
\alpha_i&=& \frac{\alpha_i^0}{1+b_i\alpha_i^0t},
 \label{sola}\\
Y_k & =&  \frac{Y_k^0u_k}{1+a_{kk}Y_k^0\int_0^t u_k},
 \label{soly}
\end{eqnarray}
where the functions $ u_k$ obey the integral system of equations
\begin{equation}
u_t=\frac{E_t}{(1+6Y_b^0\int_0^t u_b)^{1/6}}, \ \
u_b=\frac{E_b}{(1+6Y_t^0\int_0^t u_t)^{1/6} (1+4Y_\tau^0\int_0^t
u_\tau)^{1/4}}, \ \  u_\tau=\frac{E_\tau}{(1+6Y_b^0\int_0^t
u_b)^{1/2}}, \label{u}
\end{equation}
and the functions $E_k$ are given by
\begin{equation}
E_k= \prod_{i=1}^3(1+b_i\alpha_i^0t)^{c_{ki}/b_i}. \label{e}
\end{equation}

Let us stress that eqs.(\ref{sola},\ref{soly}) give the exact
solution to eqs.(\ref{a},\ref{y}), while the $u_k$'s in
eqs.(\ref{u}), although solved formally in terms of the $E_k$'s
and $Y_k^0$'s as continued integrated fractions, should in
practice be solved iteratively. Yet the important gain here is
twofold:

 i) as shown in \cite{AM}, the convergence of the
successive iterations to the exact solution can be fully
controlled analytically in terms of the initial values of the
Yukawas, allowing in practice to obtain approximations at the
level of the percent or less after one or two iterations and

ii) the structure of the solutions is explicit enough to allow for
exact statements about some regimes of the initial conditions, as
we will see later on. Furthermore, these nice features will be
naturally passed on to the solutions for the soft SUSY breaking
parameters since the latter will be obtained from
(\ref{sola}--\ref{u}) through the method of Ref.\cite{K}.

\section{The Soft Terms and Grassmannian Expansion}

An important feature of the solution (\ref{soly},\ref{u}) is that it is
written in an analytic form with the initial conditions explicitly
present. This allows one to get the same type of solution for all
the soft terms in an iterative form.

Let us describe briefly the procedure. It has been shown\cite{GG}
that the soft terms which break supersymmetry can be introduced in
a classical Lagrangian via the so-called spurion fields. This
leads to the modification of the original couplings of a rigid
theory, they become external spurion superfields depending on
Grassmannian parameters \cite{AKK}\footnote{The resulting formulae
coincide with those of Ref.\cite{JJ} except for some minor
difference in higher loops. Since we consider only one-loop RG
equations we ignore this difference here. Similar results were
obtained also in Ref.\cite{GR}}. In the MSSM it looks like
\begin{eqnarray}
 \alpha \to \tilde{\alpha}_i & = &\alpha_i(1+m_i\eta + \bar{m}_i
\bar\eta + 2 m_i\bar{m}_i\eta\bar\eta), \label{al} \\
 Y_k \to \tilde{Y}_k &=& Y_k(1-A_k\eta -\bar{A}_k\bar \eta
+(\Sigma_k+A_k\bar{A}_k)\eta \bar\eta ), \label{yu}
\end{eqnarray}
where $m_i$ are the gaugino masses, $A_k$ are the scalar triple
couplings and $\Sigma_k$ are certain combinations of the soft
masses
 $$ \Sigma_t= \tilde{m}_{Q3}^2 + \tilde{m}_{U3}^2  +
m^2_{H2}, \ \ \ \Sigma_b= \tilde{m}_{Q3}^2 + \tilde{m}_{D3}^2  +
m^2_{H1}, \ \ \ \Sigma_\tau= \tilde{m}_{L3}^2 + \tilde{m}_{E3}^2 +
m^2_{H1}.$$
 Here $\eta=\theta^2$ and $\bar \eta=\bar \theta^2$
are the spurion  fields depending on Grassmannian parameters
$\theta_\alpha, \bar\theta_\alpha$ (${\small \alpha} = 1,2$).

It has been proven in Ref.\cite{AKK} that the singular part of
effective action, which determines the renormalization properties
of any softly broken SUSY theory, is equal to that of an unbroken
one in presence of external spurion superfields. This means that
in order to calculate it in a softly broken case one just has to
take the unbroken  one, replace the couplings according to
eqs.(\ref{al},\ref{yu}) and expand over the Grassmannian
parameters $\eta$ and $\bar \eta$.

Moreover, as it has been demonstrated in \cite{K}, the same
replacement can be done directly in RG equations in order to get
the corresponding equations for the soft terms, or even in the
solutions to these equations. In the last case one obtains the
solutions to the RG equations for the soft terms. Below we
demonstrate how this procedure works in case of the MSSM, when
combined with the solutions (\ref{sola},\ref{soly}).

\section{Analytical Solution to RG Equations for the Soft Terms}

Let us now perform the substitution (\ref{al},\ref{yu}) in
(\ref{sola}-\ref{u}) and expand over $\eta$ and $\bar \eta$. Then
the linear term in $\eta$ will give us the solution for $m_i$ and
$A_k$ and the $\eta \bar \eta$ terms the ones for $\Sigma_k$. (For
simplicity, we do not consider here CP-violating effects and take
all the soft parameters to be real valued.) The resulting exact
solutions look similar to those for the rigid couplings
(\ref{sola}--\ref{u})
\begin{eqnarray}
m_i&=& \frac{m_i^0}{1+b_i\alpha_i^0t}, \label{m} \\
 A_k &=& -e_k + \frac{A_k^0/Y_k^0 +a_{kk}\int
u_ke_k}{1/Y_k^0 +a_{kk}\int u_k},\label{A}  \\ \Sigma_k &=&
\xi_k+A_k^2+2e_kA_k -\frac{(A_k^0)^2/Y_k^0
-\Sigma_k^0/Y_k^0+a_{kk}\int u_k\xi_k}{1/Y_k^0+a_{kk}\int u_k},
\label{S}
\end{eqnarray}
where the new functions $e_k$ and $\xi_k$ have been introduced
which obey the iteration equations
\begin{eqnarray}
 e_t &=& \frac{1}{E_t}\frac{d\tilde{E}_t}{d\eta}+ \frac{A_b^0\int u_b-\int
u_be_b}{ 1/Y_b^0+6\int u_b } , \nonumber
\\
 e_b &=& \frac{1}{E_b}\frac{d\tilde{E}_b}{d\eta} +\frac{A_t^0\int
u_t-\int u_te_t}{ 1/Y_t^0+6\int u_t} + \frac{A_\tau^0\int
u_\tau-\int u_\tau e_\tau}{1/Y_\tau^0 +4\int u_\tau} , \nonumber
\\
 e_\tau &=& \frac{1}{E_\tau}\frac{d\tilde{E}_\tau}{d\eta}+3\frac{A_b^0\int u_b
 -\int u_be_b}{1/Y_b^0+6\int u_b}, \nonumber
\\
&&   \nonumber \\
 \xi_t &=&
\frac{1}{E_t}\frac{d^2\tilde{E}_t}{d\eta d\bar\eta}
 +2\frac{1}{E_t}\frac{d\tilde{E}_t}{d\eta} \frac{A_b^0\int u_b-\int
u_be_b}{1/Y_b^0+6\int u_b} + 7\left(\frac{A_b^0\int u_b-\int
u_be_b}{ 1/Y_b^0 +6\int u_b} \right)^2  \nonumber
\\
 &&-\left((\Sigma_b^0+(A_b^0)^2)\int u_b
-2A_b^0\int u_be_b +\int u_b\xi_b\right)/\left( \frac{1}{Y_b^0}
+6\int u_b \right), \nonumber \\
 \xi_b &=&
\frac{1}{E_b}\frac{d^2\tilde{E}_b}{d\eta d\bar\eta}
+2\frac{1}{E_b}\frac{d\tilde{E}_b}{d\eta}\left[ \frac{A_t^0\int
u_t-\int u_te_t}{1/Y_t^0 +6\int u_t} +\frac{A_\tau^0\int
u_\tau-\int u_\tau e_\tau}{ 1/Y_\tau^0 +4\int u_\tau}
\right]\nonumber \\ && +7\left(\frac{A_t^0\int u_t-\int u_te_t}{
1/Y_t^0 +6\int u_t} \right)^2  +5\left(\frac{A_\tau^0\int
u_\tau-\int u_\tau e_\tau}{1/Y_\tau^0 +4\int u_\tau} \right)^2
\nonumber \\ && +2\left(\frac{A_t^0\int u_t-\int u_te_t}{1/Y_t^0
+6\int u_t}\right) \left(\frac{A_\tau^0\int u_\tau-\int u_\tau
e_\tau}{ 1/Y_\tau^0 +4\int u_\tau}\right)\nonumber \\
 &&-\left((\Sigma_t^0+(A_t^0)^2)\int u_t -2A_t^0\int u_te_t
+\int u_t\xi_t\right)/\left( \frac{1}{Y_t^0}+6\int u_t \right)
\nonumber \\
 && -\left((\Sigma_\tau^0+(A_\tau^0)^2)\int u_\tau
-2A_\tau^0\int u_\tau e_\tau +\int u_\tau\xi_\tau\right)/\left(
\frac{1}{Y_\tau^0} +4\int u_\tau \right) , \nonumber \\
 \xi_\tau &=&
 \frac{1}{E_\tau}\frac{d^2\tilde{E}_\tau}{d\eta d\bar\eta}
 +6\frac{1}{E_\tau}\frac{d\tilde{E}_\tau}{d\eta} \frac{A_b^0\int u_b-\int
u_be_b}{1/Y_b^0+6\int u_b} +27\left(\frac{A_b^0\int u_b-\int
u_be_b}{1/Y_b^0+6\int u_b} \right)^2 \nonumber\\
 &&-3 \left((\Sigma_b^0+(A_b^0)^2)\int u_b-2A_b^0\int u_be_b
 +\int u_b\xi_b\right)/\left( \frac{1}{Y_b^0}+6\int u_b \right).
\label{ex}
\end{eqnarray}
where the variations of $\tilde{E}_k$ should be taken at $\eta = \bar
\eta=0$ and are given by
\begin{eqnarray}
\left.\frac{1}{E_k}\frac{d\tilde{E}_k}{d\eta}\right|_{\eta,\bar
\eta=0}&=& t\sum_{i=1}^3c_{ki}\alpha_im_i^0, \label{var1}\\
\left.\frac{1}{E_k}\frac{d^2\tilde{E}_k}{d\eta
d\bar\eta}\right|_{\eta,\bar \eta=0}&=&
t^2\left(\sum_{i=1}^3c_{ki}\alpha_im_i^0 \right)^2 +2t
\sum_{i=1}^3c_{ki}\alpha_i(m_i^0)^2 -t^2
\sum_{i=1}^3c_{ki}b_i\alpha_i^2(m_i^0)^2. \label{var2}
\end{eqnarray}
When solving eqs.(\ref{u}) and (\ref{ex}) in the $n$-th iteration
one has to substitute in the r.h.s. the $(n-1)$-th iterative
solution for all the corresponding functions.

In the particular case where $Y_b=Y_\tau =0$
eqs.(\ref{u}-\ref{ex}) give an exact and well known solutions
already in the first iteration.

Let us finally note that upon inspection of the solutions
(\ref{m}--\ref{ex}), the $A_i$'s and $\Sigma_i$'s depend respectively
linearly and quadratically on the initial conditions, as expected,
and thus can be generally cast in the form:
\begin{eqnarray}
&&A_{t, b, \tau}(t) = \sum_{j=t,b,\tau} a_j(t) A_j^0 + \sum_{k=1,2,3} b_k(t)
 m_k^0, \label{linA} \\ && \nonumber \\
&&\Sigma_{t, b, \tau}(t) = \sum_{i,j=1,2,3} c_{i j}(t) m_i^0 m_j^0 +
                           \sum_{i,j=t,b,\tau} d_{i j}(t) A_i^0 A_j^0 +
                 \sum_{i=t,b,\tau,j=1,2,3 } e_{i j}(t) A_i^0 m_j^0 \nonumber\\
                  && \nonumber \\
&& \hspace{2cm} + \ k_t(t) \Sigma_t^0 + k_b(t) \Sigma_b^0 +
k_\tau(t) \Sigma_\tau^0,
 \label{linS}
\end{eqnarray}
where the various running coefficients $a_j, b_k, c_{i j}, d_{i
j}, e_{i j}$ and $k_i$ are fully determined by our solutions and
can be seen to depend exclusively on the initial conditions of the
gauge and Yukawa couplings. In Sec.6 we will evaluate these
coefficients at the E.W. scale, using a truncation of the general
solutions.

\section{Quasi-Fixed Points and the Independence of Initial Conditions}

The solutions (\ref{sola}--\ref{u}, \ref{m}--\ref{ex}) enjoy the nice
property of exhibiting the explicit dependence on initial
conditions and one can trace this dependence all the way down to the final results.
This is of special importance for the non-universal case since one
can see which of the parameters is essential and which is  washed
out during the evolution. In particular the solution for the
Yukawa couplings exhibit the fixed point behaviour when the
initial values are large enough. More precisely, in the regime
$Y_t^0, Y_b^0, Y_\tau^0 \to \infty$ with fixed finite ratios
$Y_t^0/Y_b^0= r_1, Y_b^0/Y_\tau^0= r_2$, it is legitimate to drop
$1$ in the denominators of eqs.(\ref{soly}, \ref{u}) (see 
appendix A for a proof) in which case the exact Yukawa
solutions go to the so-called IR quasi-fixed points (IRQFP)
defined by
\begin{equation}
Y_k^{FP} =  \frac{u_k^{FP}}{a_{kk}\int u_k^{FP}} \label{fp}
\end{equation}
with
\begin{equation}
u_t^{FP}=\frac{E_t}{(\int u_b^{FP})^{1/6}}, \ \
 u_b^{FP}=\frac{E_b}{(\int u_t^{FP})^{1/6} (\int u_\tau^{FP})^{1/4}} , \ \
 u_\tau^{FP}=\frac{E_\tau}{(\int u_b^{FP})^{1/2}}
\label{ufp}
\end{equation}
extending the IRQFP \cite{H} to three Yukawa couplings. What is
worth stressing here is that both the dependence on the initial
condition for each Yukawa as well as the effect of Yukawa
non-unification, $r_1, r_2$ have completely dropped out of the
runnings. ( Note that in practice this regime is already obtained
if $Y_k^0 \geq \alpha_0^{GUT}$ , assuming here for simplicity the
unification of the gauge couplings
$\alpha_1^0=\alpha_2^0=\alpha_3^0=\alpha_0^{GUT}$.) The fact that the
ratios $r_1, r_2$ drop out implies the validity of the described
properties in any $\tan \beta$ regime.

 This in turn leads to the IRQFPs for the soft
terms. Disappearance of $Y_k^0$ in the FP solution naturally leads
to the disappearance of $A_k^0$ and $\Sigma_k^0$ in the soft term
fixed points. To see this one can either go to the limit of large
$Y_k^0$ in eqs.(\ref{m}--\ref{ex}) or directly perform the
Grassmannian expansion of the FP solutions (\ref{fp},\ref{ufp}).
One gets
\begin{eqnarray}
 A_k^{FP} &=& -e_k^{FP} + \frac{\int u_k^{FP}e_k^{FP}}{\int
u_k^{FP}},\label{afp}  \\
 \Sigma_k^{FP} &=& \xi_k^{FP}+(A_k^{FP})^2+2e_k^{FP}A_k^{FP}
 -\frac{\int u_k^{FP}\xi_k^{FP}}{\int u_k^{FP}} \nonumber \\
 &= &\xi_k^{FP}
 -\frac{\int u_k^{FP}\xi_k^{FP}}{\int u_k^{FP}}-(e_k^{FP})^2+
\left(\frac{\int u_k^{FP}e_k^{FP}}{\int u_k^{FP}} \right)^2
\label{sfp}
\end{eqnarray}
with
\begin{eqnarray}
 e_t^{FP} &=& \frac{1}{E_t}\frac{d\tilde{E}_t}{d\eta}-\frac 16 \frac{\int
u_b^{FP}e_b^{FP}}{\int u_b^{FP}} ,\nonumber\\
 e_b^{FP} &=& \frac{1}{E_b}\frac{d\tilde{E}_b}{d\eta}-\frac 16 \frac{\int
u_t^{FP}e_t^{FP}}{\int u_t^{FP}} -\frac 14 \frac{\int
u_\tau^{FP}e_\tau^{FP}}{\int u_\tau^{FP}} ,\nonumber\\
 e_\tau^{FP}&=& \frac{1}{E_\tau}\frac{d\tilde{E}_\tau}{d\eta} -\frac 12 \frac{\int
u_b^{FP}e_b^{FP}}{\int u_b^{FP}} ,\nonumber\\
 && \label{exfp} \\
 \xi_t^{FP} &=&
\frac{1}{E_t}\frac{d^2\tilde{E}_t}{d\eta d\bar\eta} +\frac{7}{36}
 \left(\frac{\int u_b^{FP}e_b^{FP}}{\int u_b^{FP}}\right)^2
 -\frac 13 \frac{1}{E_t}\frac{d\tilde{E}_t}{d\eta}\frac{\int
u_b^{FP}e_b^{FP}}{\int u_b^{FP}}
 -\frac 16 \frac{\int u_b^{FP}\xi_b^{FP}}{\int u_b^{FP}}, \nonumber\\
 \xi_b^{FP} &=&
\frac{1}{E_b}\frac{d^2\tilde{E}_b}{d\eta d\bar\eta} +\frac{7}{36}
\left(\frac{\int u_t^{FP}e_t^{FP}}{\int u_t^{FP}} \right)^2
+\frac{5}{16} \left(\frac{\int u_\tau^{FP}e_\tau^{FP}}{\int
u_\tau^{FP}} \right)^2 \nonumber \\
&&-\frac{1}{E_b}\frac{d\tilde{E}_b}{d\eta}\left( \frac
13\frac{\int u_t^{FP}e_t^{FP}}{\int u_t^{FP}}+\frac 12  \frac{\int
u_\tau^{FP}e_\tau^{FP}}{\int u_\tau^{FP}} \right)
 -\frac 16 \frac{\int u_t^{FP}\xi_t^{FP}}{\int u_t^{FP}} -\frac 14
 \frac{\int u_\tau^{FP}\xi_\tau^{FP}}{\int u_\tau^{FP}},\nonumber \\
 \xi_\tau^{FP} &=&  \frac{1}{E_\tau}\frac{d^2\tilde{E}_\tau}{d\eta d\bar\eta}
  +\frac 34 \left(\frac{\int u_b^{FP}e_b^{FP}}{\int u_b^{FP}}\right)^2
  -\frac{1}{E_\tau}\frac{d\tilde{E}_\tau}{d\eta} \frac{\int
u_b^{FP}e_b^{FP}}{\int u_b^{FP}}
  -\frac 12 \frac{\int u_b^{FP}\xi_b^{FP}}{\int u_b^{FP}}.\nonumber
\end{eqnarray}
One can see from eqs.(\ref{afp},\ref{sfp}) that the constant terms
in $e_k$ and $\xi_k$ do not contribute to $A_k$ and $\Sigma_k$ and
can be dropped from eqs.(\ref{exfp}). Thus, all the dependence on
the  initial conditions $Y_k^0, A_k^0$ and $\Sigma_k^0$ disappears
from the fixed point solutions. The only dependence left is on the
gaugino masses. This is a general screening property valid for the
exact solution as well as for any of its approximate (truncated)
forms.

In view of the above screening properties of the initial
conditions at the quasi-fixed point, one should recall the
existing connection between the true IR attractive fixed point of
the Yukawa couplings and that of the soft parameters \cite{IRFP}.
What we have established can be seen as an extension of such
connections to the transient regime of quasi-fixed point at the
one-loop level.
It is also worth stressing that the above properties are
valid for any renormalization scale and are thus operative despite the
uncertainty in the choice of the physical scales.
\section{Numerical Analysis. The Role of Non-universality}

We perform now a numerical study of our solutions  in the first
and next iterations in order to demonstrate the convergence of the
procedure and to show the role of certain initial conditions. In
particular we will show that assuming that the soft terms are of
the same order of magnitude at the GUT scale the only essential
parameter is the gluino mass. All the other soft terms are
suppressed either due to the fixed point behaviour mentioned
above, or due to the smallness of $\alpha_1$ and $\alpha_2$
compared to $\alpha_3$ at low energy.

\subsection{The first iteration}

To have a solution in the first iteration one takes
eqs.(\ref{u},\ref{ex}) where in the r.h.s. the functions $u_k,
e_k$ and $\xi_k$ are taken in the $0$-th iteration. They are
\begin{equation}
u_k =E_k, \ \ e_k=\frac{1}{E_k}\frac{d\tilde{E}_k}{d\eta}, \ \
\xi_k=\frac{1}{E_k}\frac{d^2\tilde{E}_k}{d\eta d\bar\eta},
\end{equation}
where the variations of $E_k$ are given by
eqs.(\ref{var1},\ref{var2}).

For comparison we consider two particular cases:

I) $Y_t^0=5 \alpha_0, \ \ Y_b^0=Y_\tau^0 =0$,

II) $Y_t^0=Y_b^0=Y_\tau^0 = (1\div 10) \alpha_0$, ( with $\alpha_0
\approx 0.00329$).

The first case corresponds to the so-called low $\tan\beta$
regime, where the first iteration is already exact, and the second
case refers to the high $\tan\beta$ regime for an $SO(10)$-like
initial conditions.

In the first case the only essential soft terms besides the
gaugino masses are $A_t$ and $\Sigma_t$. All the soft masses of
the third generation are expressed through $\Sigma_t$ in a simple
way\cite{CK}. $A_t$ and $\Sigma_t$ have a dimension of a mass and
mass squared, respectively, and may be expanded over the initial
conditions in the following fashion (eqs.(\ref{linA}, \ref{linS}),
\begin{eqnarray}
A_t&=& 0.0359 \ A_{t 0} - 0.0167 \ m_{1 0} - 0.1598 \ m_{2 0} -
1.6284 \ m_{3 0}, \label{An} \\
 \Sigma_t &=& -0.0346 \ A_{t 0}^2  + 0.0044 \ A_{t 0} m_{1 0}
  +0.0270 \ A_{t 0} m_{2 0} + 0.1225 \ A_{t 0} m_{3 0} \nonumber \\
  && + 0.0184 \ m_{1 0}^2 -0.0057 \ m_{1 0} m_{2 0}
+ 0.2566 \ m_{2 0}^2 \nonumber  \\ &&  - 0.0335 \ m_{1 0} m_{3 0}
- 0.2735 \ m_{2 0} m_{3 0} + 6.3695 \ m_{3 0}^2 \nonumber \\  && +
  0.0359 \ (m_{H_2}^0)^2 + 0.0359 \ (\tilde{m}_{Q_3}^0)^2 +
0.0359 \ (\tilde{m}_{U_3}^0)^2, \label{Sn}
\end{eqnarray}
where the numbers are calculated for $t=\log M_{GUT}^2/M_Z^2 \approx
66$.

One can see from eqs.(\ref{An},\ref{Sn}) that the prevailing term
is that of $m_3^0$. $A_t^0$ and  $\Sigma_t^0$ decouple due to the
fixed point behaviour as  explained above and the contribution of
$m_1^0$ and $m_2^0$ is small due to smallness of $\alpha_1$ and
$\alpha_2$ compared to $\alpha_3$ at $t=66$.

The second case looks similar. We first consider the triple
couplings $A_k$. In this case one has a set of initial values $\{
A_k^0, m_i^0 \}$. In fig.1 we show the variation of the
coefficients $a_i, b_k$ of eq.(\ref{linA}) for $A_t, A_b$ and $A_\tau$ as
functions of $Y_k^0$ in the interval $Y_k^0 = \alpha_0 \div 10
\alpha_0$


\begin{figure}[t]
\epsfxsize=5.in \epsfysize=7.in \epsffile[160 0 860 700]{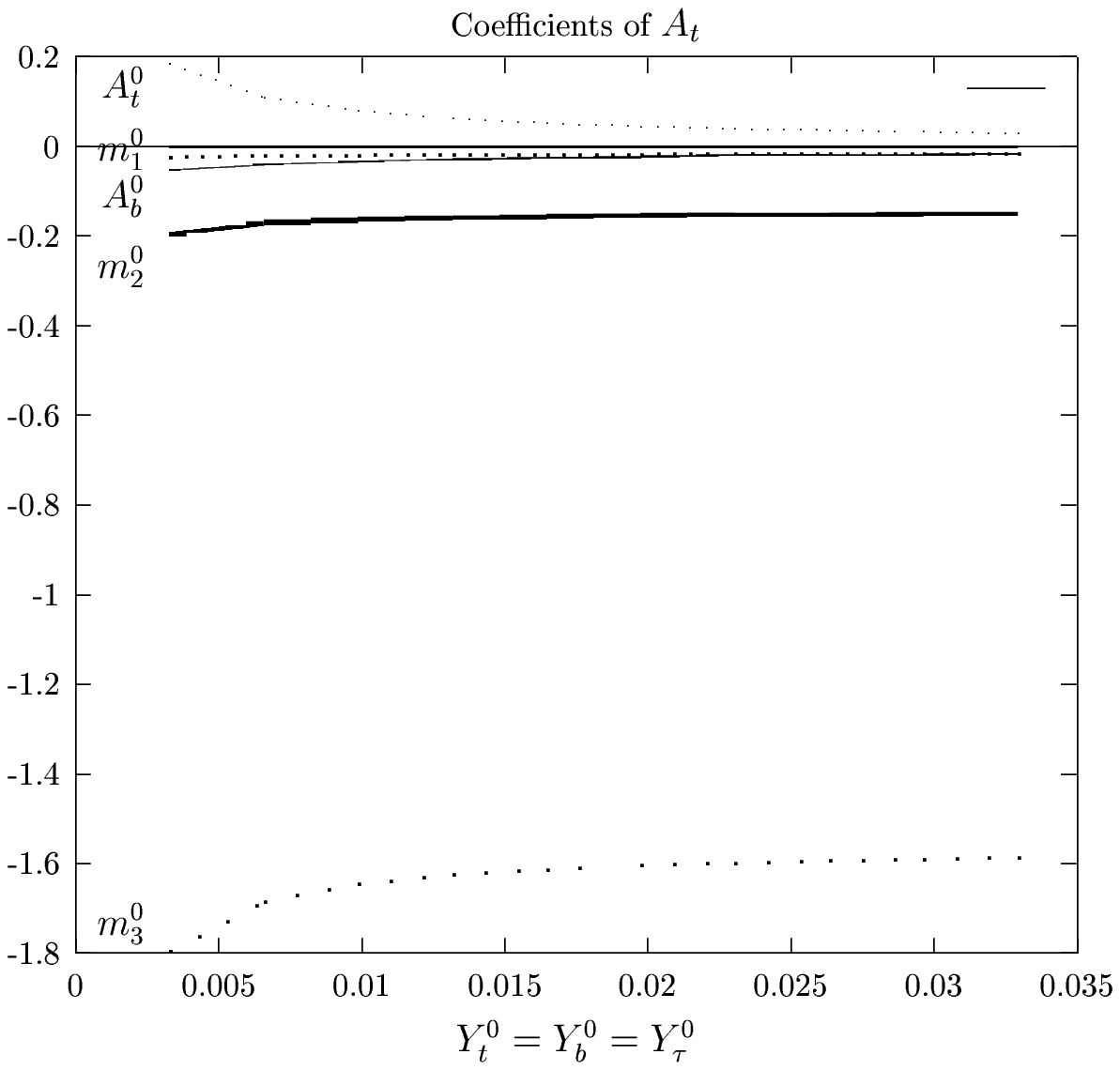}
\vspace{-17.7cm} \epsfxsize=5.in \epsfysize=7.in \epsffile[-225 0
475 700]{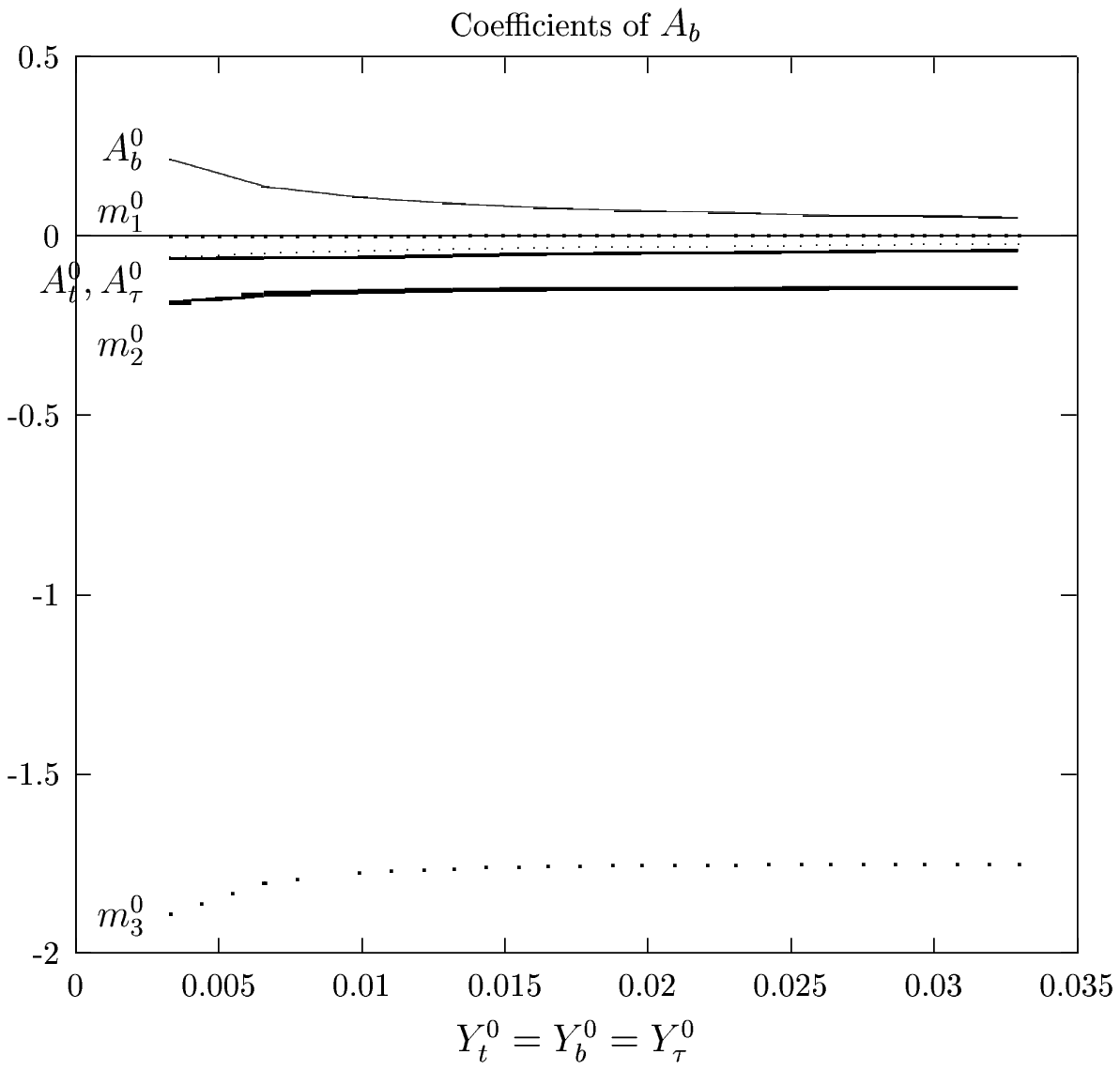} \vspace{-9.cm} \epsfxsize=5.in \epsfysize=7.in
\epsffile[0 0 700 700]{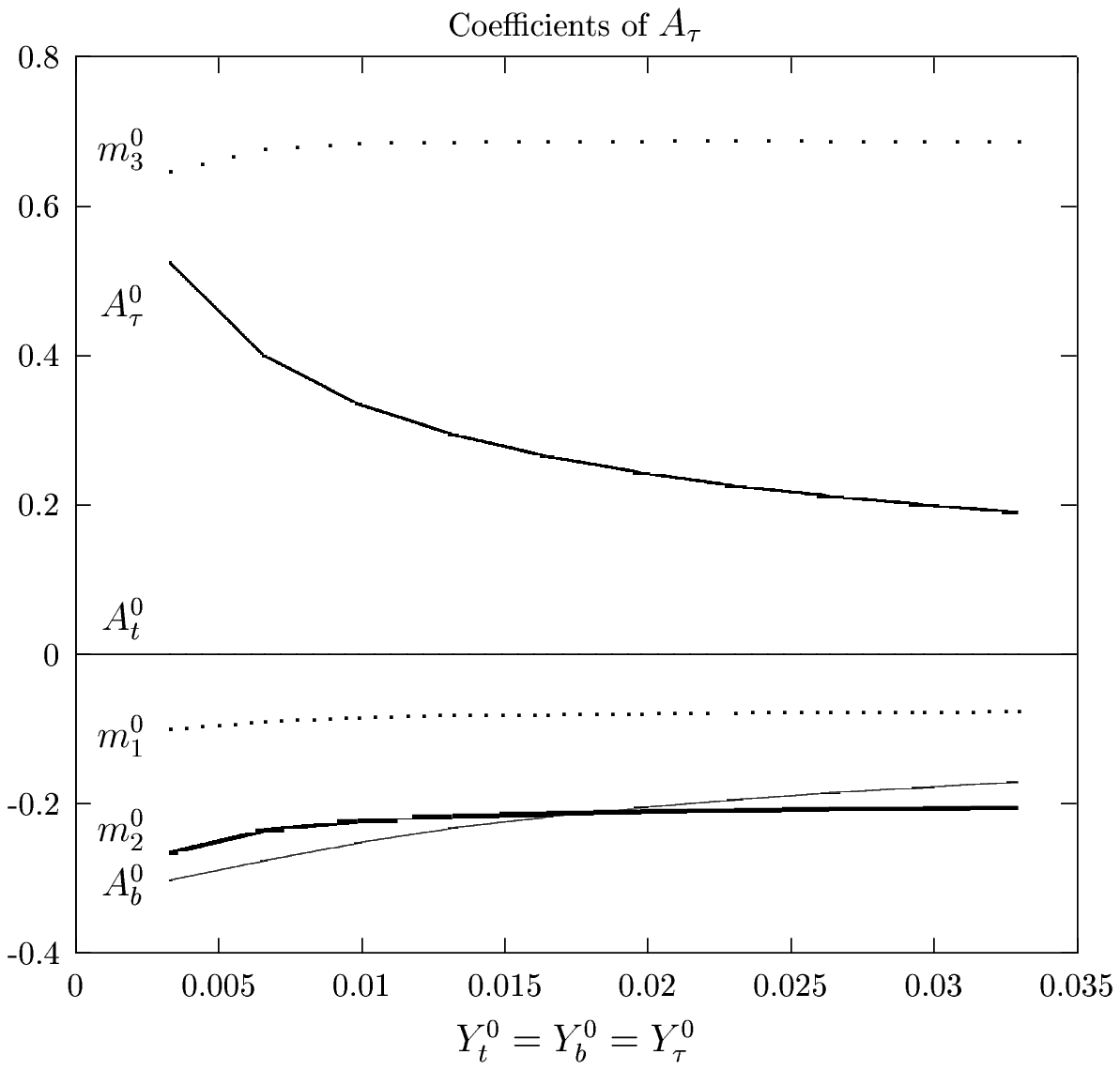} \vspace{-9cm} \caption{ Initial
value contributions of the various soft SUSY breaking parameters
to the running $A_t, A_b$ et $A_\tau$ at the EW scale, as a
function of a common initial value for the three Yukawa couplings}
\end{figure}

One can clearly see that the coefficients of $A_k^0$ are small and
have a fast decrease with increasing $Y_k^0$. The coefficients of
$m_i^0$ quickly saturate and  approach their asymptotic values
with the hierarchy $1:10:100$ for $m_1^0, m_2^0$ and  $m_3^0$. The
effect is less pronounced for $A_\tau$ due to the absence of the
SU(3) coupling in the lepton sector.

Now we come to $\Sigma_k$. We have chosen the intermediate value
of $Y_k^0=5 \alpha_0$ where the effective fixed point is practically already
reached and calculate the coefficients at $t=66$ as in
eqs.(\ref{An},\ref{Sn}).  One has
\begin{eqnarray}
 \Sigma_t &=& - 0.0497 \ A_{t 0}^2+ 0.0076 \ A_{b 0}^2  + 0.0142 \ A_{t 0} A_{b 0}
+0.0057 \ A_{t 0} m_{1 0} \nonumber \\ &&- 0.0018 \ A_{b 0} m_{1
0}+ 0.0333 \ A_{t 0} m_{2 0} - 0.0114 \ A_{b 0} m_{2 0}+
 0.1509 \ A_{t 0} m_{3 0}\nonumber  \\ && - 0.0516 \ A_{b 0}
m_{3 0}  + 0.0198 \ m_{1 0}^2 + 0.2509 \ m_{2 0}^2   + 6.3299 \
m_{3 0}^2 \nonumber \\ && - 0.0057 \ m_{1 0} m_{2 0}     -
 0.0336 \ m_{1 0} m_{3 0} - 0.2669 \ m_{2 0} m_{3 0}
- 0.0252 \ (\tilde{m}_{D_3}^0)^2 \nonumber \\ && -
 0.0252 \ (m_{H_1}^0)^2 + 0.0525 \ (m_{H_2}^0)^2 + 0.0273 \ (\tilde{m}_{Q_3}^0)^2 +
0.0525 \ (\tilde{m}_{U_3}^0)^2, \label{Stn} \\ &&  \nonumber \\
  \Sigma_b &=& + 0.0079 \ A_{t 0}^2 -0.0717 \ A_{b 0}^2 + 0.0058 \ A_{\tau 0}^2  +
0.0200 \ A_{t 0} A_{b 0}\nonumber \\ && - 0.0062 \ A_{t 0} A_{\tau
0}+ 0.0262 \ A_{b 0} A_{\tau 0} -  0.0005 \ A_{t 0} m_{1 0}  +
0.0013 \ A_{b 0} m_{1 0}\nonumber \\ && + 0.0004 \ A_{\tau 0} m_{1
0}- 0.0119 \ A_{t 0} m_{2 0} + 0.0395 \ A_{b 0} m_{2 0} - 0.0121 \
A_{\tau 0} m_{2 0}\nonumber \\ && - 0.0618 \ A_{t 0} m_{3 0} +
0.2008 \ A_{b 0} m_{3 0} - 0.0965 \ A_{\tau 0} m_{3 0}- 0.0052 \
m_{1 0}^2 \nonumber \\ && + 0.2359 \ m_{2 0}^2+ 6.8165 \ m_{3 0}^2
- 0.0027 \ m_{1 0} m_{2 0}    - 0.0037 \ m_{1 0} m_{3 0} \nonumber
\\ &&  - 0.2498 \ m_{2 0} m_{3 0}     + 0.0778 \ (\tilde{m}_{D_3}^0)^2
- 0.0502 \ (\tilde{m}_{E_3}^0)^2 + 0.0276 \ (m_{H_1}^0)^2
\nonumber \\ &&  - 0.0324 \ (m_{H_2}^0)^2 - 0.0502 \
(\tilde{m}_{L_3}^0)^2 + 0.0454 \ (\tilde{m}_{Q_3}^0)^2 - 0.0325 \
(\tilde{m}_{U_3}^0)^2, \label{Sbn}
\\ &&   \nonumber \\
 \Sigma_\tau &=& -0.0010 \ A_{b 0}^2 - 0.1947 \ A_{\tau 0}^2 + 0.1397 \ A_{b 0}  A_{\tau 0}
 -0.0195 \ A_{b 0} m_{1 0} \nonumber \\ &&  + 0.0339 \ A_{\tau 0} m_{1 0}-
 0.0409 \ A_{b 0} m_{2 0} + 0.0718 \ A_{\tau 0} m_{2 0} + 0.0894 \ A_{b 0} m_{3 0}\nonumber \\
&&  - 0.0842 \ A_{\tau 0} m_{3 0} + 0.0993 m_{1 0}^2  + 0.3527 \
m_{2 0}^2  - 2.8162 \ m_{3 0}^2 \nonumber \\ &&   - 0.0133
 \ m_{1 0} m_{2 0} +  0.0116 \ m_{1 0} m_{3 0} - 0.0687 \ m_{2 0} m_{3 0} - 0.2177 \
(\tilde{m}_{D_3}^0)^2 \nonumber \\ && +  0.2649 \
(\tilde{m}_{E_3}^0)^2
 + 0.0473 \ (m_{H_1}^0)^2 + 0.2649 \ (\tilde{m}_{L_3}^0)^2 - 0.2177 \
 (\tilde{m}_{Q_3}^0)^2.
 \label{Staun}
\end{eqnarray}
One can again see how the coefficients of the initial values of
$A_k^0$ and $\Sigma_k^0$ almost vanish and the prevailing one is
that of $(m_3^0)^2$. The next-to-leading ones are those of
$(m_2^0)^2$ and $m_2^0m_3^0$ being however almost $30$ times
smaller. This is true for both $\Sigma_t^0$ and $\Sigma_b^0$
but is less manifest for $\Sigma_\tau^0$. We note here that a
soft gaugino mass hierarchy like the one predicted by anomaly-mediated
susy breaking, $m^0_3: m^0_2:m^0_1 = 3: 0.3 :1$ \cite{Non}, enforces
even more the insensitivity of the running $A_i$'s and $\Sigma_i$'s to
the non-universality of the gaugino sector.

\subsection{The next iterations}

To demonstrate the validity of the iterative procedure and reliability of the
the first iteration we consider the effect of the next ones on the above mentioned
coefficients. We have performed the numerical integration up to the 6-th
iteration and have observed fast convergence of the coefficients to their
exact values. To show the numbers we have chosen the leading coefficients
of $m_{03}$ in $A_k$ and $m_{03}^2$ in $\Sigma_k$. In case $Y_t^0=Y_b^0=Y_\tau^0=5\alpha_0$
and $\alpha_0=0.00329$ the results are the following:
$$\begin{array}{lcccccc}
 \mbox{Iteration} &  A_t & A_b & A_\tau & \Sigma_t & \Sigma_b & \Sigma_\tau  \\
 1^{st} &  - 1.6127& - 1.7584&  0.6871&  6.3299 &6.8166 & -2.8162\\
 2^{nd} & - 1.6161&- 1.7330&  0.5037&  6.3270 & 6.6822 & -2.0989\\
 3^{rd} & - 1.6125& - 1.7372&  0.5526& 6.3192 & 6.7069 & -2.2937\\
 4^{th} &- 1.6133& - 1.7375&  0.5440&  6.3213 & 6.7054 & -2.2588\\
 5^{th} & - 1.6131 & - 1.7373& 0.5456&  6.3206 & 6.7053 & -2.2660\\
 6^{th} &  - 1.6131&  - 1.7374& 0.5454& 6.3207 & 6.7054 & -2.2649
\end{array}$$
One can see explicitly the fast convergence of the iterations.  As
expected it is worse for $A_\tau$ and $\Sigma_\tau$, so in this
case one has to take few more iterations.  We present  the general
arguments for the convergence of iterations for the soft terms in
appendix B. The advantage of this solution is that one can improve
the precision taking further iterations and in principle can
achieve any desirable accuracy.  Typically one has an accuracy of
a few percent after  2-3 iterations. This is in contrast with the
approximate solutions presented in Ref.\cite{CK} which give simple
explicit expressions but without improvement.

Taking the sixth iteration in eqs.(\ref{u},\ref{ex}) expressions for
the soft terms now look like
\begin{eqnarray*}
A_t&=& 0.0558 \ A_{t 0}-0.0294\ A_{b 0}+0.0080\ A_{\tau 0}- 0.0186
\ m_{1 0} - 0.1586 \ m_{2 0} - 1.6131 \ m_{3 0},  \\ A_b&=&
-0.0341 \ A_{t 0}+0.0984\ A_{b 0}-0.0450\ A_{\tau 0}- 0.0014 \
m_{1 0} - 0.1583 \ m_{2 0} - 1.7374 \ m_{3 0},  \\ A_\tau&=&
0.0394 \ A_{t 0}-0.2221\ A_{b 0}+0.2871\ A_{\tau 0}- 0.0825 \ m_{1
0} - 0.2344 \ m_{2 0} +0.5454 \ m_{3 0}, \\ &&\\
 \Sigma_t &=&  - 0.0487 \ A_{t 0}^2+0.0068 \ A_{b 0}^2 +0.0014 \ A_{\tau 0}^2
  + 0.0130 \ A_{t 0} A_{b 0}\\ &&- 0.0017 \ A_{t 0} A_{\tau 0} - 0.0017\ A_{b 0} A_{\tau 0}
+0.0058 \ A_{t 0} m_{1 0} - 0.0015 \ A_{b 0} m_{1 0}\\ && -0.0001
\ A_{\tau 0} m_{1 0} +  0.0335 \ A_{t 0} m_{2 0} - 0.0097 \ A_{b
0} m_{2 0} + 0.0005\ A_{\tau 0} m_{2 0} \\ && +
 0.1514 \ A_{t 0} m_{3 0} - 0.0460 \ A_{b 0} m_{3 0}+ 0.0070 \ A_{\tau 0} m_{3 0}
 + 0.0211 \ m_{1 0}^2 \\ && +
0.2547 \ m_{2 0}^2+ 6.3207 \ m_{3 0}^2- 0.0057 \ m_{1 0} m_{2 0}-
0.0340 \ m_{1 0} m_{3 0}\\ &&  - 0.2720 \ m_{2 0} m_{3 0} - 0.0294
\ (\tilde{m}_{D_3}^0)^2-
 0.0214 \ (m_{H_1}^0)^2 + 0.0558 \ (m_{H_2}^0)^2 \\ &&  + 0.0264 \ (\tilde{m}_{Q_3}^0)^2 +
0.0558 \ (\tilde{m}_{U_3}^0)^2+ 0.0080 \ (\tilde{m}_{L_3}^0)^2  +
0.0080 \ (\tilde{m}_{E_3}^0)^2, \\ &&  \nonumber \\
 \Sigma_b&=& +0.0067 \ A_{t 0}^2-0.0736 \ A_{b 0}^2 -0.0035 \ A_{\tau 0}^2
 + 0.0200 \ A_{t 0} A_{b 0}\\ && - 0.0064 \ A_{t 0} A_{\tau 0}+0.0302\ A_{b 0} A_{\tau 0}
- 0.0003 \ A_{t 0} m_{1 0} - 0.0000 \ A_{b 0} m_{1 0} \\&& +0.0021
\ A_{\tau 0} m_{1 0}-0.0110 \ A_{t 0} m_{2 0}+0.0396 \ A_{b 0}
m_{2 0}-0.0076\ A_{\tau 0} m_{2 0}\\ && -0.0613 \ A_{t 0} m_{3 0}
+0.2356 \ A_{b 0} m_{3 0}-0.0945 \ A_{\tau 0} m_{3 0} - 0.0013 \
m_{1 0}^2 \\ &&
 + 0.2532 \ m_{2 0}^2+ 6.7053 \ m_{3 0}^2 - 0.0030 \ m_{1 0} m_{2 0}
 - 0.0053 \ m_{1 0} m_{3 0} \\ && - 0.2578 \ m_{2 0} m_{3 0}
+0.0984 \ (\tilde{m}_{D_3}^0)^2  +
 0.0534 \ (m_{H_1}^0)^2 -0.0341 \ (m_{H_2}^0)^2 \\ &&  + 0.0642 \ (\tilde{m}_{Q_3}^0)^2
-0.0341 \ (\tilde{m}_{U_3}^0)^2- 0.0450 \ (\tilde{m}_{L_3}^0)^2  -
0.0450 \ (\tilde{m}_{E_3}^0)^2, \\ && \\
 \Sigma_\tau&=& +0.0009 \ A_{t 0}^2-0.0106 \ A_{b 0}^2 -0.1862 \ A_{\tau 0}^2
 + 0.0077 \ A_{t 0} A_{b 0}\\ && - 0.0146 \ A_{t 0} A_{\tau 0}+0.1334\ A_{b 0} A_{\tau 0}
+0.0021 \ A_{t 0} m_{1 0} - 0.0190 \ A_{b 0} m_{1 0}\\ && + 0.0350
\ A_{\tau 0} m_{1 0} +0.0007 \ A_{t 0} m_{2 0}-0.0350 \ A_{b 0}
m_{2 0}+0.0743\ A_{\tau 0} m_{2 0}\\ &&  -0.0397 \ A_{t 0} m_{3 0}
+0.1288 \ A_{b 0} m_{3 0}-0.1090 \ A_{\tau 0} m_{3 0} +0.1029 \
m_{1 0}^2 \\&&
 + 0.3907 \ m_{2 0}^2 -2.2649 \ m_{3 0}^2-0.0140 \ m_{1 0} m_{2 0}
 +0.0160 \ m_{1 0} m_{3 0}\\ && - 0.0504 \ m_{2 0} m_{3 0}
-0.2221 \ (\tilde{m}_{D_3}^0)^2  +
 0.0650 \ (m_{H_1}^0)^2 +0.0394 \ (m_{H_2}^0)^2 \\ &&  -0.1827 \ (\tilde{m}_{Q_3}^0)^2
+0.0394 \ (\tilde{m}_{U_3}^0)^2+0.2871 \ (\tilde{m}_{L_3}^0)^2  +
0.2871 \ (\tilde{m}_{E_3}^0)^2
\end{eqnarray*}
to be compared with eqs.(\ref{Stn},\ref{Sbn},\ref{Staun}). These
numbers can be now used for the calculation of the soft parameters
and masses using eqs.(\ref{mass}) with arbitrary initial
conditions for the soft terms. Being calculated for
$Y_t^0=Y_b^0=Y_\tau^0=5\alpha_0$ and $\alpha_0=0.00329$ they are
practically independent of the initial values of Yukawa couplings
provided the latter are big enough.
 
\section{Towards the Physical Masses}

The values of $\Sigma_k$ completely
define those of the soft masses for squarks, sleptons and Higgses
due to linear relations which follow from the RG equations
\cite{CK} and read, after relaxing the universality assumption\footnote{
Note that even though a trace term ``$Tr (Y_{hypercharge} m^2)$'' is generically
present in the RGE in the non-universal case, it cancels out in eq.(\ref{mass})},
\begin{eqnarray}
\tilde{m}^2_{Q3}&=&(\tilde{m}^0_{Q3})^2+\frac{128f_3+87f_2-11f_1}{122}
+\frac{17(\Sigma_t-\Sigma_t^0)+20(\Sigma_b-\Sigma_b^0)
-5(\Sigma_\tau-\Sigma_\tau^0)}{122},\label{mass} \\
\tilde{m}^2_{U3}&=&(\tilde{m}^0_{U3})^2+\frac{144f_3-108f_2+144/5f_1}{122}
+\frac{42(\Sigma_t-\Sigma_t^0)-8(\Sigma_b-\Sigma_b^0)
+2(\Sigma_\tau-\Sigma_\tau^0)}{122},\nonumber\\
\tilde{m}^2_{D3}&=&(\tilde{m}^0_{D3})^2+\frac{112f_3-84f_2+112/5f_1}{122}
+\frac{-8(\Sigma_t-\Sigma_t^0)+48(\Sigma_b-\Sigma_b^0)
-12(\Sigma_\tau-\Sigma_\tau^0)}{122},\nonumber\\
{m}^2_{H1}&=&({m}^0_{H1})^2+\frac{-240f_3-3f_2-57/5f_1}{122}
+\frac{-9(\Sigma_t-\Sigma_t^0)+54(\Sigma_b-\Sigma_b^0)
+17(\Sigma_\tau-\Sigma_\tau^0)}{122},\nonumber\\
{m}^2_{H2}&=&({m}^0_{H2})^2+\frac{-272f_3+21f_2-89/5f_1}{122}
+\frac{63(\Sigma_t-\Sigma_t^0)-12(\Sigma_b-\Sigma_b^0)
+3(\Sigma_\tau-\Sigma_\tau^0)}{122},\nonumber\\
\tilde{m}^2_{L3}&=&(\tilde{m}^0_{L3})^2+\frac{80f_3+123f_2-103/5f_1}{122}
+\frac{3(\Sigma_t-\Sigma_t^0)-18(\Sigma_b-\Sigma_b^0)
+35(\Sigma_\tau-\Sigma_\tau^0)}{122},\nonumber\\
\tilde{m}^2_{E3}&=&(\tilde{m}^0_{E3})^2+\frac{160f_3-120f_2+32f_1}{122}
+\frac{6(\Sigma_t-\Sigma_t^0)-36(\Sigma_b-\Sigma_b^0)
+70(\Sigma_\tau-\Sigma_\tau^0)}{122},\nonumber
\end{eqnarray}
where
$$f_i=\frac{(m_i^0)^2}{b_i}\left(1-\frac{1}{(1+b_i\alpha_0t)^2}\right).$$

At this level one can already make rough qualitative statements about
the physical scalar masses. We note first that,
as can be seen from the above equations,  the sensitivity to
the initial conditions reappears partly in the running of the soft
scalar masses, even in the vicinity of the IRQFP. However, this dependence
remains confined in the initial values of the soft masses themselves in a
universal (scale independent) form, and in the initial conditions
of the gaugino soft masses through the $f_i$'s and the $\Sigma$'s.
[The dependence on the initial values of Yukawa couplings as well as on the $A^0$'s and the 
$\Sigma^0$'s, that could come from the running, remain completely screened.]
The ratios giving the universal sensitivity of the 
running soft scalar masses to the soft scalar masses initial conditions
is as follows:

$$\begin{array}{lccccccc}
\! &(\tilde{m}^0_{Q3})^2 \! : &(\tilde{m}^0_{U3})^2 \! : &(\tilde{m}^0_{D3})^2 \! : &({m}^0_{H1})^2 \! : &({m}^0_{H2})^2 \! : &(\tilde{m}^0_{L3})^2 \! : &(\tilde{m}^0_{E3})^2 \\
(\tilde{m}_{Q3})^2 \! &  17     \!  &   -3.4   \!  &   -4     \!  &    -3       \!  & -3.4     \!  &   1        \!  &    1 \\
(\tilde{m}_{U3})^2 \! & -17 \! & 40\! & 4\! &  3  \! &  -21 \! &  -1\! & -1 \\
(\tilde{m}_{D3})^2 \! &-5\! & 1\! & 9.25\! & -4.5\! &  1\! & 1.5\! & 1.5 \\
({m}_{H1})^2       \!  &-5\! & 1\! & -6\! & 5.67\! & 1\! &  -1.89 \! &  -1.89 \\ 
({m}_{H2})^2 &-17\! &  -21\! & 4\! & 3\! & 19.67 \! &  -1\! &  -1\\
(\tilde{m}_{L3})^2 \! & 5 \! & -1\! & 6\! & -5.67\! & -1\! &  29\! &  -11.67 \\
(\tilde{m}_{E3})^2 \! & 5 \! & -1\! & 6\! &  -5.67\! &  -1\! & -11.67 \! & 8.67  
\end{array}$$
These numbers are renormalization scale independent and give the trend
of the relative sensitivity in the vicinity of the IRQFP.

On the other hand, the dependence on the initial soft gaugino masses is
renormalization scale dependent. At the electroweak scale ($ t \simeq 66$),
one finds that in the soft masses of the third squark
generation and of the Higgs doublets the sensitivity to $(m_3^0)^2$ remains 
leading (by a factor of 15 to 25) as compared to $(m_2^0)^2$. 
In contrast, a large cancellation occurs for the sleptons, leading to 
comparable sensitivities to $(m_3^0)^2$ and $(m_2^0)^2$ in $\tilde{m}_{E3}$, 
and even a bigger sensitivity to $(m_2^0)^2$ (by a factor of 4) in 
$\tilde{m}_{L3}$.\\

To go further to the physical scalar masses, one has to consider the behaviour
of the  $\mu$ parameter which enters the mixing of the left and right states.
The running of this parameter has the simple form
$ \mu(t) \sim \mu_0  exp[ \int_0^t (\alpha - Y)] $ where $\alpha, Y$ are generic
 gauge and Yukawa couplings. Thus, here too, the initial conditions for the
 Yukawas are screened near the IRQFP in the evolution of $\mu$,
the $A_i$'s and $\Sigma_i$'s being absent anyway. However, when the
electroweak symmetry breaking (EWSB) is required to take place
radiatively, the $\mu$ parameter becomes, as usual, correlated to
the other parameters of the MSSM at the electroweak scale. To be
specific, in the leading one-loop top/stop-bottom/sbottom
approximation to the EWSB conditions, the sensitivity of $\mu$ to
the initial conditions will come basically from the soft scalar
masses of the Higgs doublets and scalar partners of the third
quark generation \cite{YJK}. As stated before, the latter
dependence is dominated, on one hand  by the initial conditions of
the soft scalar masses, in a well determined scale independent
way, and on the other hand by the (scale-dependent) $m^0_3$
contributions. The same dependence pattern is then taken over to
the physical scalar masses. A further inclusion of the scalar
$\tau$ contributions to the EWSB conditions will basically not
affect this dependence pattern. Indeed, although
$\tilde{m}^2_{L3}$ and $\tilde{m}^2_{E3}$ have comparable sensitivity
between $m^0_2$ and $m^0_3$ at the electroweak scale, they are
less sensitive to this sector altogether than to the squark soft masses. 
All in all our analytical results allow to draw at this
stage a qualitative
 sensitivity hierarchy for the {\sl physical} scalar masses:\\

-- basically no sensitivity to Yukawa couplings initial conditions
(whether unified or not), or $A_i^0$ initial conditions (whether
universal or not),

-- important sensitivity to initial conditions of the soft gaugino
masses, however basically only through $m^0_3$, {\sl i.e.} weak
sensitivity to non-universality of this sector,

 -- important sensitivity to initial conditions of the soft scalar
 masses, however through a universal scale independent pattern.


\section{Conclusion}

In the present paper we have obtained general analytical forms for
the solutions of the one-loop renormalization group equations in
the top/bottom/$\tau$ sector of the MSSM. These solutions are
valid for any value of $\tan \beta$ as well as any non-universal
initial conditions for the soft SUSY breaking parameters and
non-unification of the Yukawa couplings. They allow  a general study
of the evolution of the various parameters of the MSSM and to trace back,
 sector-wise, the sensitivity to initial conditions of the Yukawa couplings and the
soft susy breaking parameters. We have established
analytically a generic screening  of non-universality, in the
vicinity of the infrared quasi fixed points. In practice, this property gives 
the
general trend of the behaviour, despite the large number of free parameters,
and even when one is not very close to such a quasi fixed point.
This shows that non-universality of the $A$ parameters
and gaugino soft masses, as well as Yukawa unification conditions, would
basically have no influence on the squark and Higgs spectra. The main
input from the gaugino sector comes from the soft gluino mass contribution
 (which dominates by far the other two), {\sl i.e.} insensitive to non-universality
conditions of this sector. The only substantial sensitivity to non-universality
is associated to the initial conditions of the scalar soft masses, but
is renormalization scale independent and well defined. 
A similar pattern holds for the sleptons, apart from the fact that now the
contribution of the wino soft mass becomes comparable to that of the gluino,
yet the overall sensitivity to the gaugino sector is much smaller than in the
case of the squarks.

Detailed illustrations of the physical spectrum, 
including the lightest Higgs, will be given in a subsequent study.

\vspace{0.5cm}

{\bf Acknowledgments}

\vspace{0.3cm}

 We are grateful to members of
``GDR-supersym\'etrie"  for useful discussions and to S.Codoban
for suggesting to us useful tricks for the iterative numerical integration.
D.K. would like to thank the University of Montpellier II for hospitality and
CNRS for financial support.
 \newpage
\renewcommand{\theequation}{A.\arabic{equation}}
\setcounter{equation}{0}
\section*{Appendix A: Screening of initial conditions and of non-universality
 at the IRQFP}
We give here a short proof that in the regime $Y_t^0, Y_b^0,
Y_\tau^0 \to \infty$ with fixed finite ratios $Y_t^0/Y_b^0= r_1,
Y_b^0/Y_\tau^0= r_2$ the Yukawas become insensitive both to
$Y_k^0$ and $r_1, r_2$. This result would be immediate from
eq.(\ref{soly}) by dropping $1$ in the denominator were it not for
the fact that the $u_k$'s have also a non-trivial dependence on
the large initial conditions. Let us rewrite eqs.(\ref{u}) in the
form
\begin{eqnarray}
&&\tilde{u}_t^{(n)}=\frac{\tilde{E}_t}{(1+6Y_b^0\int \tilde{u}_b^{(n-1)})^{1/6}} \nonumber \\
&&\tilde{u}_b^{(n)}=\frac{\tilde{E}_b}{(1+6Y_b^0\int \tilde{u}_t^{(n-1)})^{1/6} (1+4Y_b^0\int
\tilde{u}_\tau^{(n-1)})^{1/4}} \nonumber \\
&&\tilde{u}_\tau^{(n)}=\frac{\tilde{E}_\tau}{(1+6Y_b^0\int
\tilde{u}_b^{(n-1)})^{1/2}}, \label{uFP}
\end{eqnarray}
with $$ \tilde{u}_k^{(0)} \equiv \tilde{E}_k, \ \ (k=t, b, \tau),
$$
 where the twiddled quantities are obtained from the non twiddled
ones by proper rescaling with $r_1$ or $r_2$, and we indicate
explicitly the order of iteration. It is now easy to show
inductively that if at the $n^{th}$ iteration
\begin{equation}
 \tilde{u}_k^{(n)} \sim \frac{(\tilde{u}_k^{(n)})^{FP}}{(Y_b^0)^{p_k}} \ \
\mbox{for  } \ Y_b^0 \to \infty, \ \ \mbox{with} \ 0 <p_k < 1 \ ,
\label{prop}
\end{equation}
where $(\tilde{u}_k^{(n)})^{FP}$ is $Y_b^0$ independent but $r_1, r_2$ dependent,
then the same is true at the $(n+1)^{th}$ iteration. Furthermore,
 since (\ref{prop}) is obviously true for $n=1$ as can be easily seen from
(\ref{uFP}) we conclude that the exact $u_k$'s behave also like
$$
 \tilde{u}_k \sim \frac{{(\tilde{u}_k)}^{FP}}{{(Y_b^0)}^{p_k}} \
 \mbox{with} \ 0 <p_k < 1 .
$$
 This means that the $1$'s can be legitimately dropped both in
eq.(\ref{soly}) and eq.(\ref{u}). The complete cancellation of
$Y_b^0, r_1$ and $r_2$ in the final result is then obvious,
leading to eqs.(\ref{fp}, \ref{ufp}).

\section*{Appendix B: Convergence of iterations for the soft
terms}
\renewcommand{\theequation}{B.\arabic{equation}}
\setcounter{equation}{0}
In this appendix we prove that the convergence of the $e_i's$ and
$\xi_i's$ is automatic once that of the $u_i's$ is achieved. In particular
this means that a controllable behaviour is expected whatever the
initial conditions for the soft parameters may be.

Let us define
\begin{equation}
{\cal E}(t)= \left(
\begin{array}{ccc}
0  &e_t(t)& 0 \\
e_b(t) &0 &0 \\
0 &e_{\tau}(t)& 0 \\
\end{array} \right)
\label{mate}
\end{equation}
\begin{equation}
{\cal U}(t_1; t)= \left(
\begin{array}{ccc}
0  &  - {\cal U}_b(t_1;t)& 0 \\
- {\cal U}_t(t_1;t) &0 &- {\cal U}_{\tau}(t_1;t) \\
0 & - 3 {\cal U}_b(t_1;t)& 0 \\
\end{array} \right)
\label{matu}
\end{equation}where
\begin{equation}
  {\cal U}_i(t_1;t) \equiv \frac{u_i(t_1)}{1/Y_i^0 +
             a_i \int_0^t u_i} \label{matu1}, \ i=t, b, \tau
\end{equation}
and $a_t=a_b=6, a_\tau=4$,

\begin{equation}
{\cal C}(t)= \left(
\begin{array}{ccc}
\frac{\displaystyle 1}{\displaystyle E_t}\frac{\displaystyle
d\tilde{E}_t}{\displaystyle d\eta} + A_b^0 \int_0^t {\cal
U}_b(t_1;t) & 0 & 0 \\ 0 & \frac{\displaystyle 1}{\displaystyle
E_b}\frac{\displaystyle d\tilde{E}_b}{\displaystyle d\eta} +
\sum_{k=t, \tau}
 A_k^0 \int_0^t {\cal U}_k(t_1;t) & 0 \\
0 & 0& \frac{\displaystyle 1}{\displaystyle
E_\tau}\frac{\displaystyle d\tilde{E}_\tau}{\displaystyle d\eta} +
3 A_b^0 \int_0^t {\cal U}_b(t_1;t) \\
\end{array} \right)
\label{matC}
\end{equation}
with $C(0) =0$. The system of integral equations for the $e_i$'s can then be
written in the matrix form
\begin{equation}
{\cal E}(t) = {\cal C}(t) + \int_0^t {\cal U}(t_1;t) {\cal E}(t_1) dt_1
\label{veceq}
\end{equation}

To prove the convergence of ${\cal E}(t)$ we the define the mapping
$ {\cal E} \to {\cal E}'$:
\begin{equation}
{\cal E}'(t) = {\cal C}(t) + \int_0^t {\cal U}(t_1;t) {\cal E}(t_1) dt_1
\label{mapeq}
\end{equation}
and the norm $\parallel . \parallel$ through

\begin{equation}
\parallel M(t) \parallel \equiv \sup_{0 \leq t \leq T} \{ \max \mid M_{i j}(t)\mid\}
\label{normeq}
\end{equation}
for any matrix $M$ in a given evolution interval $[ 0, T]$.
One then has the inequality

\begin{equation}
\mid \int_0^t ({\cal U E})_{i j} \mid\leq  (\sum_k \int_0^t \mid {\cal U}_{i k}\mid )
 \parallel {\cal E} \parallel \label{ineq1}
\end{equation}
valid for any $i, j$. On the other hand, one has from eqs.(\ref{matu}, \ref{matu1})

\begin{equation}
\sum_k \int_0^t \mid {\cal U}_{i k}\mid = \left \{
\begin{array}{cc} \frac{ \int u_b}{ 1/Y_b^0 + 6 \int
u_b} \leq \frac 16 & (i=1) \\ \frac{ \int u_t}{ 1/Y_t^0 + 6 \int
u_t} +\frac{ \int u_\tau}{ 1/Y_\tau^0 + 4 \int u_\tau} \leq
\frac{5}{12} & (i=2) \\ 3 \frac{ \int u_b}{ 1/Y_b^0 + 6 \int u_b}
\leq \frac 12 & (i=3) \\
\end{array} \right. \label{ineq2}
\end{equation}
Combining the above inequalities (\ref{ineq1}, \ref{ineq2}) with eq.(\ref{mapeq})
 one obtains
\begin{equation}
\parallel {\cal E}_1' - {\cal E}_2' \parallel \leq \frac 12 \parallel
 {\cal E}_1 - {\cal E}_2 \parallel
\end{equation}
that is, the mapping (\ref{mapeq}) is a contraction, the solution to eq.(\ref{veceq})
 is unique
and approximated at worse with an error of $1/2^n$, after
$n$ iterations. Actually, the situation is much better than given by
this upper bound error, as one can see from the numerical illustrations
of section {\bf 6.2}. Finally, we note that the rational is exactly the
same for the convergence of the $\xi$'s. Indeed, apart from a different definition
for ${\cal C}(t)$,  the $\xi$'s satisfy a matrix
equation similar to (\ref{veceq}) with the same ${\cal U}$
as the one given in (\ref{matu}).


\end{document}